\documentclass{ws-ijmpe}

\begin{document}

\markboth{Thorsten Renk and J\"{o}rg Ruppert}{Prospects of Medium Tomography using 2-, 3- and 4-particle Correlations for a Semi-Hard Trigger}

\catchline{}{}{}{}{}

\title{PROSPECTS OF MEDIUM TOMOGRAPHY USING 2-, 3- AND 4-PARTICLE CORRELATIONS FOR A (SEMI)-HARD TRIGGER}

\author{\footnotesize THORSTEN RENK}

\address{Department of Physics, PO Box 35 FIN-40014 University of Jyv\"askyl\"a, Finland and\\ Helsinki Institute of Physics, PO Box 64 FIN-00014, University of Helsinki, Finland\\
trenk@phys.jyu.fi}

\author{J\"{O}RG RUPPERT}

\address{Department of Physics, McGill University, 3600 Rue University, \\Montreal, QC, Canada, H3A 2T8\\
ruppert@physics.mcgill.ca}

\maketitle

\begin{history}
\end{history}

\begin{abstract}
Hard partons propagating through hot and dense matter lose energy, leading to the observed depletion of hard hadron spectra in nucleus nucleus collision as compared to scaled proton proton collisions. This lost energy has to be redistributed in the medium due to the conservation of energy, which is manifest in the $p_T$ dependence of the angular correlation pattern of hadrons associate with a (semi-) hard trigger. While at low $p_T$ a splitting of a broad peak is observed, at high $p_T$ the structure shows vacuum width, albeit with reduced yield. This sugests a transfer of energy from hard partons to a collectively recoiling medium.
We present a systematic study of these phenomena using a realistic medium evolution and a Monte-Carlo simulation of the experimental trigger and show what information about the medium can be derived from multiparticle correlations.
\end{abstract}

\section{Introduction}

Energy loss of a high $p_T$ 'hard' parton travelling through low $p_T$ 'soft' matter has long been 
recognized as a promising tool to study the initial high-density phases of ultrarelativistic
heavy-ion collisions (URHIC) \cite{Jet1,Jet2,Jet3,Jet4,Jet5,Jet6}. However, if one considers the whole dynamical system created in the collision of two relativistic nuclei and not only the partons emerging from a particular hard scattering vertex, energy is not lost but rather redistributed into the medium. 

Measurements of angular correlations of hadrons associated with a given hard trigger allow, as a function of associate hadron momentum, to study how and at what scales this redistribution of energy and momentum takes place. Such measurements for semi-hard hadrons with 1 GeV $< p_T < $ 2.5 GeV associated with a trigger 2.5 GeV $< p_T <$ 4.0 GeV have shown a surprising splitting of the away side peak for all centralities but peripheral collisions, qualitatively very different from a broadened away side peak observed in p-p or d-Au collisions \cite{PHENIX-2pc}. On the other hand, for a high $p_T$ trigger and associate hadron back to back jet peaks with vacuum width albeit reduced strength have been observed \cite{Dijets1,Dijets2}.

As most promising explanation for these findings, the assumption that Mach shockwaves are excited by the energy lost from the hard parton to the medium has been brought forward \cite{Stoecker,Shuryak}. While we focus on colorless sound here \cite{Stoecker,Shuryak}, we point out that also colored collective modes could contributed in a deconfined phase by a supersonically traveling jet, if the plasma's collective longitudinal mode exhibit a space-like dispersion relation \cite{Ruppert1,Ruppert2}. For a short overview of those as well as alternative mechanisms, see \cite{Ruppert2} and references therein.

In the following, we investigate under what conditions signals from such shockwaves remain observable in the dynamical environment of a heavy-ion collision, provided that a realistic description of the evolving medium and the experimental trigger conditions are taken into account, and what properties of the medium are reflected in different momentum regimes. Here, we summarize and expand on \cite{Mach1} and \cite{Mach2}.

\section{The model}

We perform a Monte Carlo (MC) simulation of hard back to back processes in a medium in order to calculate the induced hard hadronic correlation pattern. The model is described in detail in \cite{Mach1,Mach2,Correlations}. There are four main elements: 1) the primary hard pQCD process 2) the description of the soft medium and 3) the energy loss from hard to soft degrees of freedom and 4) the simulation of shockwave propagation and its modification of soft medium decoupling.

For the soft medium we use the parametrized evolution model defined in \cite{Parametrized} which gives a good description of bulk matter transverse momentum spectra and HBT correlation radii.
The energy loss for a given parton path inside this medium is described in a probabilistic language.
In order to determine the probability  $P(\Delta E,  E)_{\rm path}$ for a 
hard parton with energy $E$ to lose the energy $\Delta E$ while traversing the medium on its 
trajectory, we make use of a scaling law \cite{JetScaling} which allows to relate the dynamical 
scenario to a static equivalent one for each trajectory $\xi(\tau)$ by calculating 
 
\begin{equation}
\label{E-omega}
\omega_c({\bf r_0}, \phi) = \int_0^\infty d \xi \xi \hat{q}(\xi)
\quad {\rm and} \quad
\langle\hat{q}L\rangle ({\bf r_0}, \phi) = \int_0^\infty d \xi \hat{q}(\xi)
\end{equation} 
 
as a function of the jet production vertex ${\bf r_0}$ and its angular orientation $\phi$. We set 
$\hat{q} \equiv 0$ whenever the decoupling temperature of the medium $T = T_F$ is reached.
Using the numerical results of \cite{QuenchingWeights}, we obtain $P(\Delta E)_{\rm path}$ 
for $\omega_c$ and $R=2\omega_c^2/\langle\hat{q}L\rangle$ 
as a function of jet production vertex and the angle $\phi$.

We generate hard pQCD back to back events inside the evolving bulk matter model, using Eq.~(\ref{E-omega}) to determine $P(\Delta E)_{\rm path}$ for each parton, followed by leading and next to leading fragmentation once the parton emerges from the medium. The simulation describes the yield of hard back to back correlations well \cite{Correlations}. Selecting events leading to a trigger determines the distribution of vertices and parton energies which we use as a starting point to simulate the away side shockwave. 
First, we 
determine the direction of the away side parton in the transverse plane and rapidity. In order to
take into account intrinsic $k_T$, we calculate a random $k_T$ kick by sampling a Gaussian distribution with $\approx 1~{\rm GeV}$ width. 
We have verified that this distribution, folded with the width of the near
side peak reproduces the width of the away side peak in the case of d-Au and 60-90\% peripheral Au-Au collisions.

Since we are not interested in folding the result with a steeply falling spectrum but rather into the
energy deposited on average in a given volume element we focus on the average energy loss of the away side parton
$\langle \Delta E \rangle = \int_0^\infty P(\Delta E) \Delta E d\Delta E$ in the following.
We assume that a fraction $f$ of the energy and momentum lost to the medium 
excites a shockwave characterized by a dispersion relation $E = c_s p$ while a fraction
$(1-f)$ in essence heats the medium and induces collective drift along the jet
axis conserving longitudinal momentum.
We calculate the speed of sound $c_s$ locally from a quasiparticle description of the equation of state as
measured on the lattice as $c_s^2 = \partial p(T)/ \partial \epsilon(T)$.

The dispersion relation along with the energy and momentum deposition determines the initial angle of propagation
of the shock front with the jet axis (the 'Mach angle') as $\phi = \arccos c_s$. We discretize the time into small intervals $\Delta \tau$, calculate the energy deposited in that time as
$E(\tau) = \Delta \tau \cdot dE/d\tau$. We then propagate the part of the shockfront remaining in the midrapidity
slice (i.e. in the detector acceptance). Each piece of the front is propagated with the local speed of sound and the angle of propagation is
constantly corrected as 
\begin{equation}
\label{E-c_s}
\phi = \arccos \frac{\int_{\tau_E}^\tau c_s(\tau) d\tau }{(\tau - \tau_E)}
\end{equation}
where $c_s(\tau)$ is determined by the propagation path.

Once an element of the wavefront reaches the freeze-out condition $T = T_F$, a hydrodynamical
mode cannot propagate further. We assume at this point that the energy contained in the shockwave
is not used to produce hadrons but rather is converted into kinetic energy. In the local
restframe, we then have a matching condition for the dispersion relations
\begin{equation}
E = c_s p \quad \text{and} \quad E = \sqrt{M^2 + p^2} - M
\end{equation} 
where $M = V \left(p(T_F) + \epsilon(T_F)\right)$ is the 'mass' of a volume element at freeze-out temperature.
Once we have calculated the additional boost $u_\mu^{\rm shock}$ a volume element receives from the shockwave using the
matching conditions, we employ the Cooper-Frye formula
\begin{equation}
\label{E-CF}
E \frac{d^3N}{d^3p} =\frac{g}{(2\pi)^3} \int d\sigma_\mu p^\mu
\exp\left[\frac{p^\mu (u_\mu^{flow} + u_\mu^{shock}) - \mu_i}{T_f}\right]
\end{equation}
to convert the fluid element into a hadronic distribution.
The resulting momentum spectrum is thus a thermal two component spectrum resulting from an integration involving
volume not part of the shockwave and volume receiving an additional boost from the shockwave.

\section{Flow}

\begin{figure}[htb]
\epsfig{file=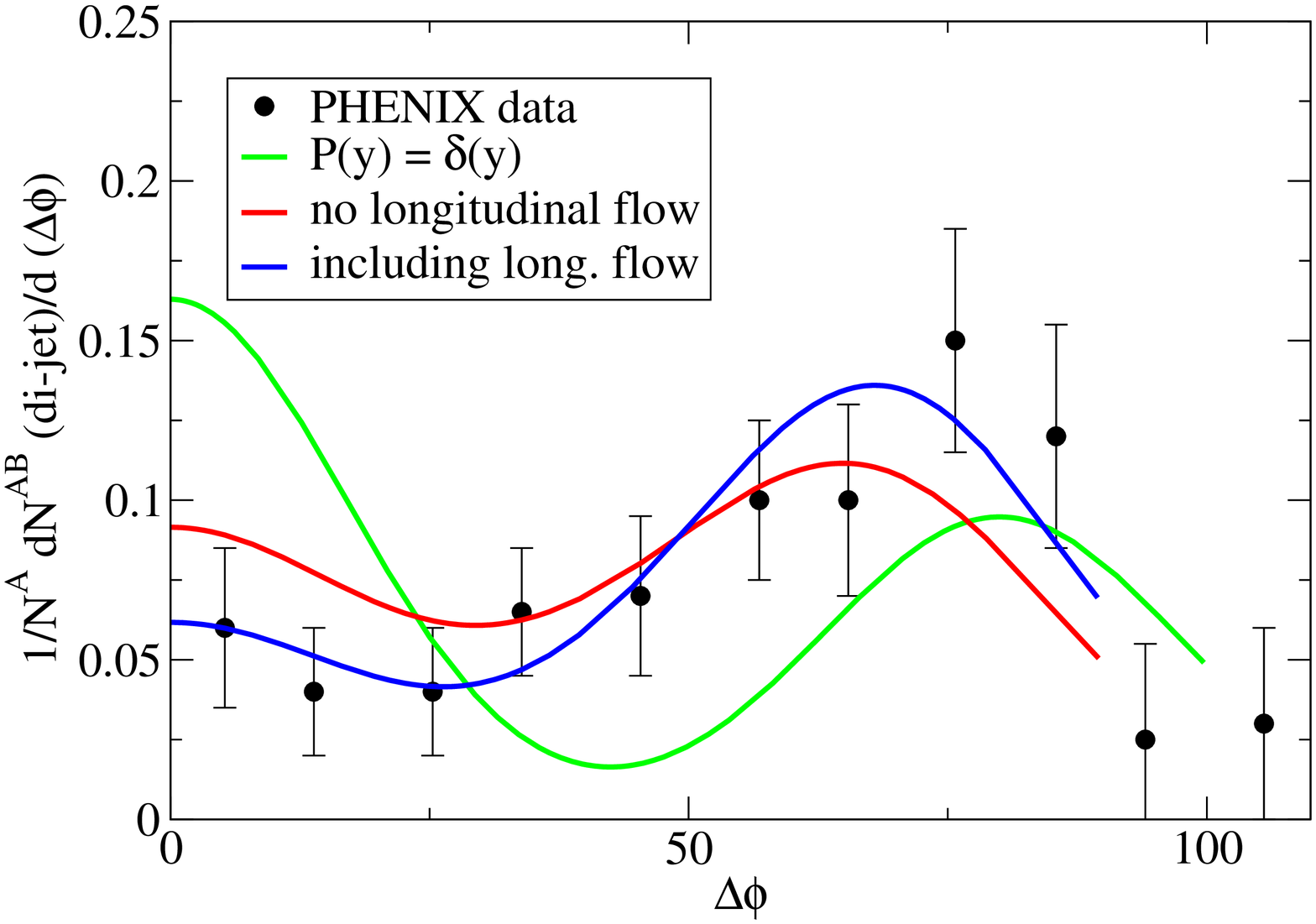, width=6.2cm}\epsfig{file=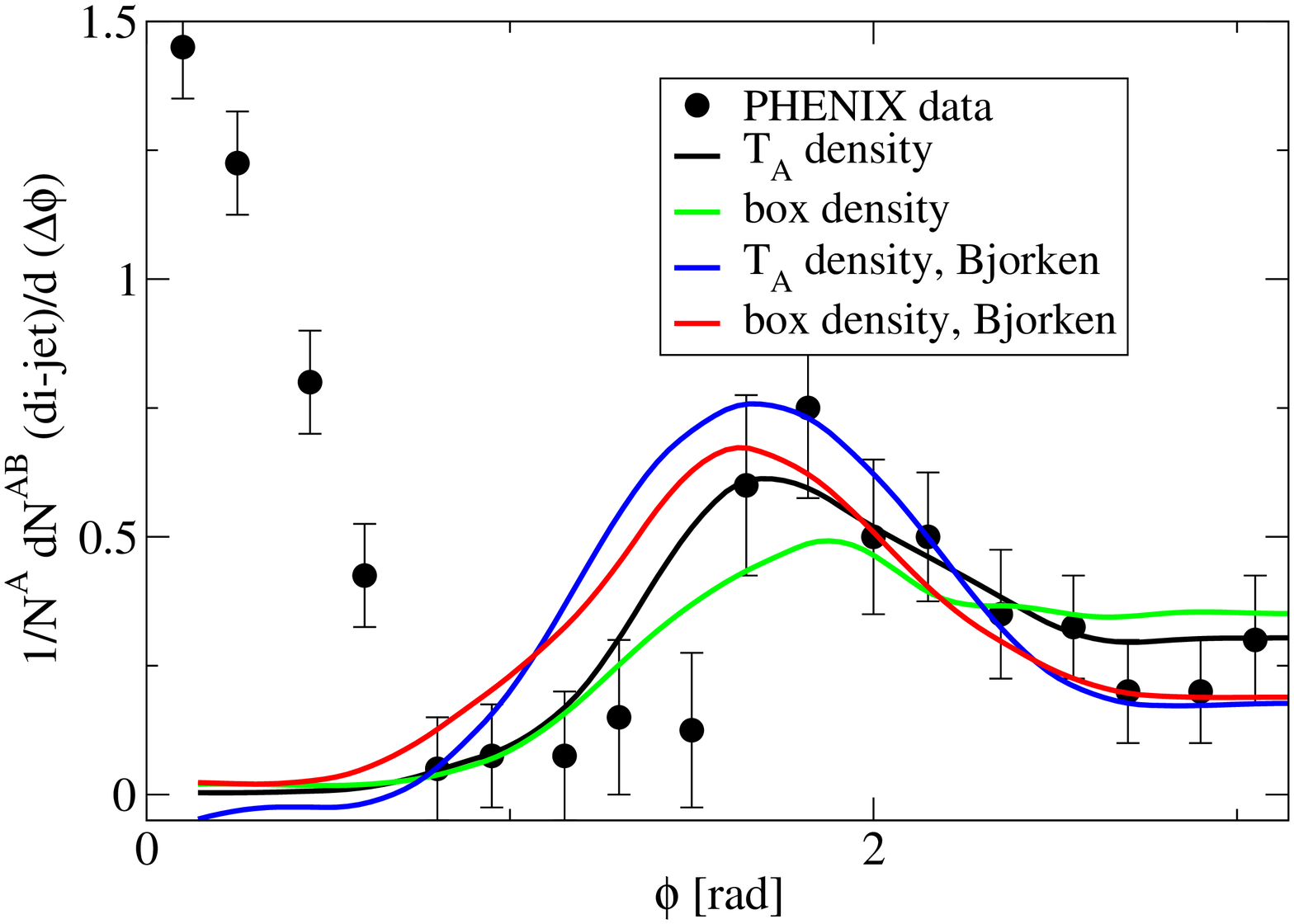, width=6.2cm}
\caption{\label{F-1}Left panel: Calculated 2-particle correlation under the assumption that a) the away side parton is always at midrapidity and the excited mode doesn't couple to flow (green) b) the excited mode doesn't couple to flow (red) and c) including realistic $P(y)$ and longitudinal flow effect. Right panel: Variation induced in the correlation signal by changing the soft background medium density distribution and evolution (see text, not acceptance-averaged).}
\end{figure}

The fact that the medium in which shockwaves propagate isn't static has considerable impact on the resulting shape of the medium recoil. Since the wavefront propagates with $c_s$ relative to the flowing medium, its maximal longitudinal extension has to be determined by
\begin{equation}
\label{E-LFlow}
\left.\frac{dz}{dt}\right|_{z=z(t)}=\left.\frac{u(z,R,t)+c_s(T(z,R,T))}{1+u(z,R,t)c_s(T(z,R,t))}\right|_{z=z(t)}.
\end{equation}
where $u(z,R,t)$ is the longitudinal flow velocity. The effect of this is illustrated in Fig.~\ref{F-1}, left panel. If, for a trigger at midrapidity, the away side parton would always be found at midrapidity and the shockfront would not be modified by longitudinal flow, a peak at a relatively large angle would be observed. However, averaging over the pQCD probability $P(y)$ to find the away side parton at the (unobserved) rapidity $y$ shifts the peak inward as one gets contributions from events where the away side parton is outside the acceptance. In those only a slice of the cone falls into the acceptance in which the opening angle doesn't appear maximal. 
However, an elongation of the cone in longitudinal direction is induced by longitudinal flow which lessens this effect. We include longitudinal flow using Eq.~(\ref{E-LFlow}) and determine the corresponding rapidity boost from the fireball evolution model\cite{Mach2}.   One can also turn this argument around: scenarios which do not predict a substantial elongation as it emerges in the Mach cone framework are strongly disfavored by the data since the correlation signal is easily washed out via rapidity averaging over $P(y)$ \cite{Mach2}.

Another manifestation of flow is visible in Fig.~\ref{F-1}, right panel. We point out that this 2-particle correlation signal is the one corresponding  to our recent MC simulation of 3-particle correlations (see below).
In this specific calculation we suppressed a full rapidity averaging, leading to a somewhat larger angle than expected if averaging were performed. 
We show two different longitudinal expansion patterns (Bjorken and non-Bjorken evolution) and two different transverse densities (box and nuclear profile $T_A$). Due to the  position of the freeze-out hypersurface at large radii for the box density, the shockwave gets on average longer exposure to transverse flow in this scenario. 
We find the following pattern: The observed angle is consistently larger for longitudinal Bjorken expansion. This is due to the fact that in this scenario initial cooling is rapid and therefore the average temperature quickly approaches the phase transition temperature $T_C$ where $c_s$ is small (and hence the Mach angle large, cf. Eq.~(\ref{E-c_s})). Furthermore, the peak is consistently less pronounced for box density scenarios where there is longer exposure to transverse flow. As argued in \cite{Mach1,HQproc}, transverse flow can erase a peak for associate momenta above 1 GeV if the direction of flow and shockwave propagation are not aligned.

This has important consequences for 3-particle correlations which are sensitive to the question if only one wing or both wings of the shockwave are seen for each event. We calculate 3-particle correlations as factorized two particle correlations, i.e. we calculate a 2-particle correlation as outlined above, subtract the background (which is obtained by evaluating Eq.~(\ref{E-CF}) without $u_\mu^{\rm shock}$) and assume that a second hadron in the same event has the same probability distribution of being found at a given angle with the trigger independent of the position of the first hadron. This assumption ignores the possibility of genuine 3-particle correlations. However, the only such genuine 3-particle effect which is expected is due to conservation of momentum $\sum p_L = p_{away}$ and $\sum {\bf p_T} = 0$ which is shared across $O(10-20)$ hadrons \cite{Mach1} and hence their influence on the signal is very small. We then average the generated 3-particle correlation over several hundred events, as large event by event fluctuations in the correlation signal occur due to the probabilistic nature of the collisions (modeled in our MC approach).

\begin{figure}[htb]
\epsfig{file=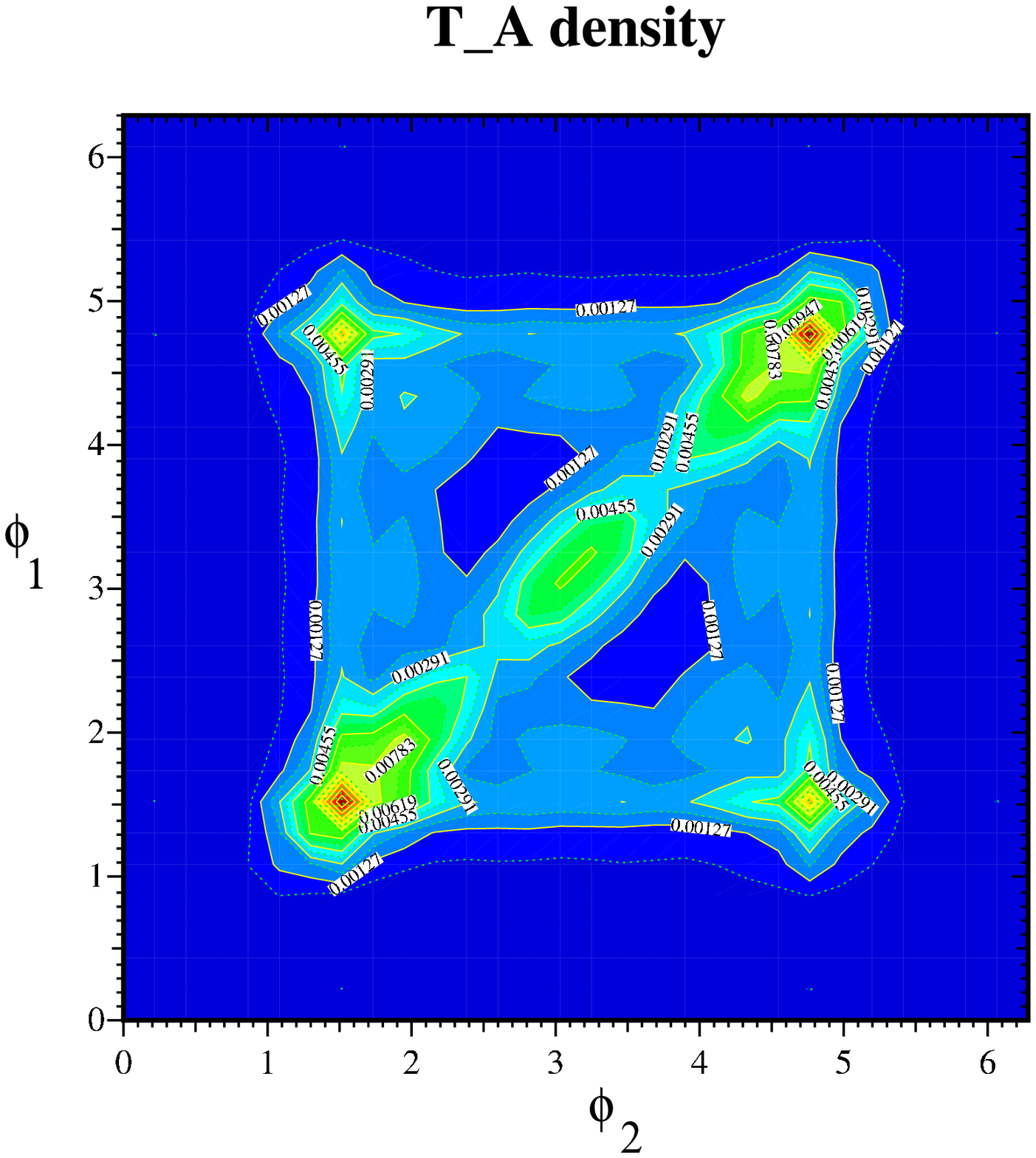, width=6.2cm} \epsfig{file=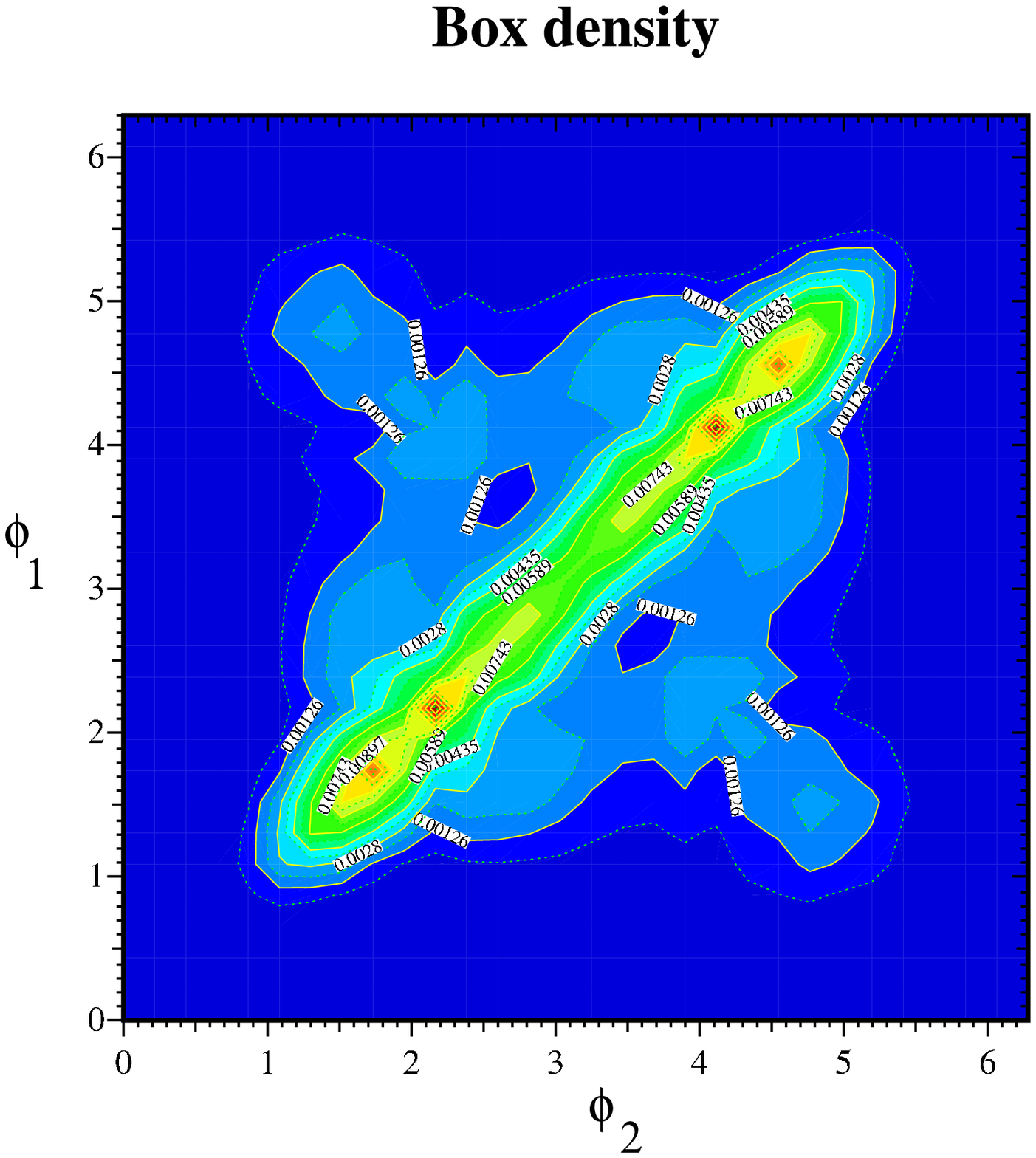, width=6.2cm}
\vspace*{-2cm}
\caption{\label{F-3pc}3-particle correlation strength for associate hadrons with $1.0 < p_T < 2.5$ GeV as a function of angles $\phi_1$ and $\phi_2$ with the trigger for a nuclear profile function shaped and a box shaped medium density distribution.}
\end{figure}

We show the resulting 3-particle correlations for two different density distributions (not including the region around the trigger) in Fig.~\ref{F-3pc}. The result fits nicely into the earlier observation that transverse flow tends to erase one (or even both wings of the shock cone). In this case, only the diagonal region is populated, whereas if both wings of the shock cone projection into the transverse plane survive in an average event, off-diagonal elements appear. Thus, we can understand that the box density where flow has greater effect shows less strength in the off-diagonal peaks, as in typical events only one wing of the shock cone is observed.

As a side remark, we note that there is no scenario involving flow in which the strength of off-diagonal peaks equals the strength of the peaks on the diagonal. The correlation pattern found in the case of the nuclear profile $T_A$ density resembles very much the pattern measured by STAR \cite{STAR-3pc}.

\section{The ridge correlation}

In addition to the rich correlation structure on the away side, there has also been an observation of a long-range $\Delta \eta$ correlation on the near side, the so-called ridge \cite{Ridge}. Theoretical explanations suggested for this calculation include e.g. longitudinal broadening of jet cones by turbulent color fields \cite{Turbulent}. 

So far, the ridge correlation has only been observed on the near side whereas the Mach cone like correlation has been seen only on the away side. While it is clear that a ridge-like correlation cannot be observed on the away side due to the unknown rapidity of the away side parton, the intriguing question remains why a cone should not be seen on the near side. While in the simulation about 75\% of all partons leading to a triggered hadron do not experience energy loss \cite{Correlations,Blackness}, there is energy transfer into the medium from the rest which should give rise to a shockwave if the above scenario is correct.

However, the timescales for development of a cone are rather different on near and away side, as the bulk of trigger partons originates from $\sim 3$ fm depth from the surface (thus the average away side parton has a pathlength $O(10)$ fm). It is thus tempting to identify the ridge with an incompletely formed shockwave and try to explain near and away side phenomenology with the same mechanism.

Clearly, this interpretation cannot be verified by exact calculations in our present model. However, one can verify that basic scales are correct and that such an interpretation quite naturally fits the general framework. Using Eq.~(\ref{E-LFlow}) for the typical near side parton vertex, we find that the timescale is sufficient for the longitudinal flow to expand the correlation into the interval $-1<y<1$. Furthermore, if the ridge correlation is driven by energy lost from the trigger parton, it should not scale with the trigger energy, but with the expected energy loss given a trigger energy, a quantity which is approximately constant in the kinematic range probed so far. Finally, if the ridge is akin to the shockwave, its main effect should be an additional boost to hadrons, i.e. an upward shift in apparent temperature with a mass ordering of this shift just like in the case of radial flow, i.e. the shift in apparent temperature should be stronger for nucleons than for pions.

The most useful measurement to discover more about the relationship between ridge and cone is to trigger on a back to back hadron pair. In this case, the rapidity position of near and away side parton is approximately known, i.e. if there is a ridge on the away side it should become visible. Furthermore, the vertex distribution of events can be dialed by chosing appropriate momenta on both sides: If equal momenta back to back are required the situation has to be symmetric and most vertices occur in the center, otherwise the vertex distribution is shifted towards the medium surface on the high-momentum side. Thus, such a 4-particle correlation measurement could also check if properties of ridge and cone depend crucially on the average in-medium pathlength of the parent parton.

\section{Discussion}

We have shown that the correlation pattern of associate hadrons back to back with a hard trigger can be understood in terms of shockwaves induced by the energy loss of th away side parton. Many properties of the observed signal find a natural explanation once this fundamental mechanism is adopted \cite{HQproc}.

If so, the resulting correlation pattern shows a sensitivity to several properties of the medium: First, the angular position of the peak is sensitive to the average temperature of the system and hence sensitive to the expansion (which determines the rate of cooling) or, if the expansion is assumed to be known, to the EOS (which determines $c_s(T)$). Flow corrections modify the position somewhat. In particular, longitudinal flow is crucial to spread the correlation signal over a large rapidity interval whereas transverse flow has a strong influence on the structure of the observed 3-particle correlations. Within the given framework, we could demonstrate that the data favour one density profile assumption over another, however more detailed studies are needed before medium properties can be inferred in a more direct way from the data.

We would like to stress that a detailed simulation of the background medium, the acceptance and trigger conditions and the energy loss as performed in our calculations is crucial for these features. Especially we observe strong event by event fluctuations of the 2- and 3-particle correlation signals, and while some events may be similar to the average, the majority is not. Hence the simulation of a single event cannot be considered as a good representative of the ensemble and should not be compared with measured data. Likewise, the average speed of sound in medium cannot reliably be obtained from the measured angle without a simulation including the effects of flow.

Finally, we have outlined a 4-particle correlation measurement which would provide valuable insights into the nature of both ridge and cone correlation.

\section*{Acknowledgements}

We would like to thank J.~Cassalderay-Solana, F.~Wang, J.~Rak, H.~St\"{o}cker, J.~Putschke and M.~van Leeuwen for valuable discussions. This work was supported by the Academy of Finland, project 206024, and by the Natural Sciences and Engineering Research Council of Canada.

\end{document}